\documentclass[aip,jcp,graphicx,reprint]{revtex4-1}
\usepackage{amsmath}
\usepackage{graphicx}
\usepackage{hyperref}
\draft 

\begin{document}

\title{Analysis of the Heyd-Scuseria-Ernzerhof density functional parameter space}

\author{Jonathan E. Moussa}
\email{godotalgorithm@gmail.com}

\author{Peter A. Schultz}
\affiliation{Sandia National Laboratories, Albuquerque, NM 87185, USA}

\author{James R. Chelikowsky}
\affiliation{Center for Computational Materials, Institute for Computational Engineering and Sciences, Departments of Physics and Chemical Engineering, University of Texas, Austin, TX  78712, USA}

\date{\today}

\begin{abstract}
The Heyd-Scuseria-Ernzerhof (HSE) density functionals are popular for their ability to improve
 the accuracy of standard semilocal functionals such as Perdew-Burke-Ernzerhof (PBE), particularly for semiconductor band gaps.
 They also have a reduced computational cost compared to hybrid functionals,
 which results from the restriction of Fock exchange calculations to small inter-electron separations.
These functionals are defined by an overall fraction of Fock exchange and a length scale for exchange screening.
We systematically examine this two-parameter space to assess the performance of hybrid screened exchange (sX) functionals
 and to determine a balance between improving accuracy and reducing the screening length,
 which can further reduce computational costs.
Three parameter choices emerge as useful: ``sX-PBE'' is an approximation to the sX-LDA screened exchange density functionals
 based on the local density approximation (LDA);
 ``HSE12'' minimizes the overall error over all tests performed; and ``HSE12s'' is a range-minimized functional that
 matches the overall accuracy of the existing HSE06 parameterization but reduces the Fock exchange length scale by half.
Analysis of the error trends over parameter space produces useful guidance for future improvement of density functionals.

\end{abstract}


\maketitle

\section{Introduction}

The exact, nonlocal form for the many-electron Fock exchange energy is known from Hartree-Fock theory.
However, it is a standard practice in density functional theory (DFT) to compute this energy
 by integrating a local energy density per electron that is specified by the local electron density and its derivatives.
One of the most popular of these semilocal density approximations
 is the Perdew-Burke-Ernzerhof (PBE) model\cite{PBE}.
Accuracy can be improved by mixing the PBE exchange energy with a fraction of the exact nonlocal Fock exchange energy,
 producing hybrid functionals such as PBE0\cite{PBE0}.
Exchange mixing can also depend on distance: the Heyd-Scuseria-Ernzerhof (HSE) density functional\cite{HSE03}
 retains only short-range Fock exchange
 and preserves the accuracy of PBE0 while avoiding
 the cost and pathologies\cite{Fock_metal} of long-range Fock exchange.
Unlike in traditional Kohn-Sham theory,
 hybrid functionals generate nonlocal potentials,
 which is described by a generalized Kohn-Sham (GKS) theory\cite{GKS}.
Ideally, an exact GKS functional can produce both a highest occupied molecular orbital (HOMO)
 and a lowest unoccupied molecular orbital (LUMO) that have orbital energies
 equivalent to the negative ionization potential (IP) and electron affinity (EA), respectively.
In practice, HSE and other short-range Fock exchange functionals are
 accurate for band gaps (IP+EA) of semiconductors\cite{SC40},
 but fail significantly for large-gap insulators\cite{PBE_eps}, 
 molecules\cite{sX_correction}, interfaces\cite{HSE_surf}, and nanostructures\cite{HSE_nano}.

The HSE functional form defines a 2-dimensional space of DFT functionals\cite{HSE03}, set by the fraction of Fock exchange, $a$,
 at zero electron separation and a length scale, $\omega^{-1}$, on which the short-range Fock exchange is computed,
\begin{align} \label{sX_xc}
 E_{\mathrm{xc}}^{\mathrm{HSE}} = &a E_{\mathrm{x}}^{\mathrm{HF,SR}}(\omega) + (1-a) E_{\mathrm{x}}^{\mathrm{PBE,SR}}(\omega)  \notag \\
 &+ E_{\mathrm{x}}^{\mathrm{PBE,LR}}(\omega) + E_{\mathrm{c}}^{\mathrm{PBE}}.
\end{align}
The short-range Fock exchange (in hartrees) is calculated using the
 spinful Kohn-Sham density matrix $\rho_{\sigma,\sigma'}(\mathbf{r},\mathbf{r}')$,
\begin{align} \label{sX}
 E_{\mathrm{x}}^{\mathrm{HF,SR}}(\omega) = -\frac{1}{2}\sum_{\sigma,\sigma'} \int d\mathbf{r} d\mathbf{r}'  & \frac{\mathrm{erfc}(\omega |\mathbf{r}-\mathbf{r}'|)}{|\mathbf{r}-\mathbf{r}'|} \notag \\ & \times |\rho_{\sigma,\sigma'}(\mathbf{r},\mathbf{r}')|^2  ,
\end{align}
 while the long-range and remaining short-range exchange are derived from the exchange-hole formulation of PBE\cite{PBE_hole}.
The intent of HSE was to achieve accuracy equivalent to PBE0,
 therefore the exchange fraction was limited to its PBE0 value of  $a = 0.25$.
The HSE06 reparameterization of the HSE form\cite{HSE06} was also based on a variation of $\omega$ and not $a$.
The remaining $(\omega,a)$ space has been explored only sparsely\cite{HSE_explore} with
 results suggesting that different choices of $\omega$ and $a$ improve the accuracy of different physical properties.

\begin{figure}
\includegraphics{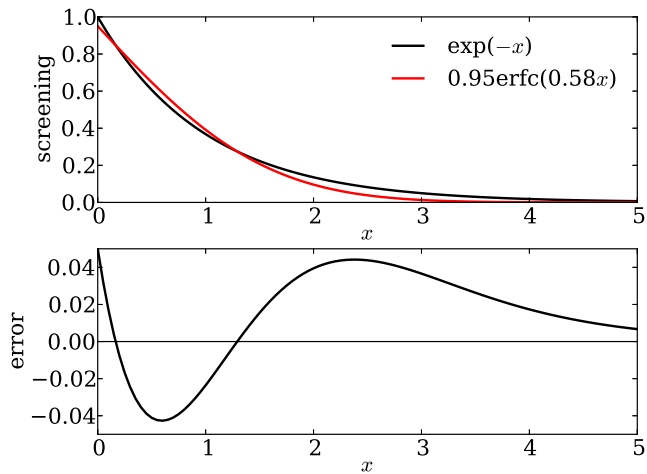}
\caption{\label{screen}Approximation of the exponential exchange screening function with a scaled complementary error function.
Both functions are displayed in the upper panel and their difference is displayed in the lower panel.}
\end{figure}

HSE is not the first density functional to combine a short-range fragment of Fock exchange with a semilocal model of long-range exchange.
The concept has appeared previously as screened exchange (sX) by Bylander and Kleinman\cite{sX_LDA},
 where a combination of short-range Fock exchange and the local density approximation (LDA) of correlation and long-range exchange,
 denoted\cite{GKS} sX-LDA, was used to improve LDA band gaps of semiconductors.
This application of short-range Fock exchange to band gap estimation is
 motivated by the Coulomb-hole-plus-screened-exchange (COHSEX) approximation to electron quasiparticle theory\cite{COHSEX}
 in combination with the Thomas-Fermi model of screening.
Based on this motivation, an $\exp(-r/r_\mathrm{TF})$ screening factor is used in place of $\mathrm{erfc}(\omega r)$ in Eq. (\ref{sX})
 and the Fock exchange fraction is set to unity ($a=1$).
The difference in screening form is insignificant, since the two variants can approximate each other,
\begin{equation} \label{screen_approx}
 \mathrm{exp}(-x) \approx  0.95 \mathrm{erfc}(0.58 x),
\end{equation}
 with a maximum pointwise error of $\approx 0.05$ (see Fig. \ref{screen}).
However, HSE uses a single $\omega$ parameter for all systems, while the $r_\mathrm{TF}$ parameter in sX-LDA
 is set to the Thomas-Fermi screening length based on average valence electron density $\overline{\rho}$ (in atomic units),
\begin{equation}\label{rtf}
 r_{\mathrm{TF}} = \frac{1}{2} \left( \frac{\pi}{3 \overline{\rho}} \right)^{1/6}.
\end{equation}
System-dependence of the Fock exchange length scale is the most important difference between HSE and sX-LDA,
 especially because $\overline{\rho}$ is not well-defined for a molecule without an arbitrary definition of molecular volume.

Given the flexibility of the HSE form, we fully explore the $(\omega,a)$ parameter space
 to determine if its accuracy and efficiency can be improved further.
Also, we assess the variation in results over this space to explain why HSE is accurate
 and reveal its unresolved shortcomings.
First, we search for a ``sX-PBE'' functional in HSE space that reproduces the performance of sX-LDA,
 which we define by setting $a = 0.95$ and optimizing $\omega$ for band gap estimation.
Despite the use of a system-independent $\omega$ instead of a system-dependent $r_\mathrm{TF}$,
 sX-PBE is able to surpass the accuracy of sX-LDA.
This can be explained by the small degree of variation in $r_\mathrm{TF}$ values.
Next, we search for a ``HSE12'' parameterization that minimizes the overall error over multiple test sets.
The result is a modest improvement over the HSE06 parameterization,
 which comes from using a larger Fock exchange fraction ($a=0.313$).
Finally, we search for a range-minimized ``HSE12s'' 
 parameterization that minimizes the value of $\omega^{-1}$ while preserving the accuracy of HSE06.
This results in reduction of $\omega^{-1}$ by a factor of two, which can reduce the cost of evaluating Eq. (\ref{sX})
 by reducing the number of $\rho$ off-diagonals that need to be computed to achieve a target accuracy.
After evaluating multiple tests sets, we observe significantly different behavior in two quantities:
 molecular IPs and EAs as approximated by HOMO and LUMO energies.
We can explain this phenomenon as a fundamental limitation of the HSE functional form that
 results from the use of an environment-independent and homogeneous screening of exchange
 that is unable to account for inhomogeneous screening environments.

\section{Computational details}

The analysis we present in this paper is based on DFT calculations of isolated molecules and periodic solids.
All molecular calculations are performed in \textsc{gaussian 09}
\footnote{ \textsc{gaussian 09}, Revision B.01,
 M. J. Frisch, G. W. Trucks, H. B. Schlegel, G. E. Scuseria, 
 M. A. Robb, J. R. Cheeseman, G. Scalmani, V. Barone, B. Mennucci, 
 G. A. Petersson, H. Nakatsuji, M. Caricato, X. Li, H. P. Hratchian, 
 A. F. Izmaylov, J. Bloino, G. Zheng, J. L. Sonnenberg, M. Hada, 
 M. Ehara, K. Toyota, R. Fukuda, J. Hasegawa, M. Ishida, T. Nakajima, 
 Y. Honda, O. Kitao, H. Nakai, T. Vreven, J. A. Montgomery, Jr., 
 J. E. Peralta, F. Ogliaro, M. Bearpark, J. J. Heyd, E. Brothers, 
 K. N. Kudin, V. N. Staroverov, T. Keith, R. Kobayashi, J. Normand, 
 K. Raghavachari, A. Rendell, J. C. Burant, S. S. Iyengar, J. Tomasi, 
 M. Cossi, N. Rega, J. M. Millam, M. Klene, J. E. Knox, J. B. Cross, 
 V. Bakken, C. Adamo, J. Jaramillo, R. Gomperts, R. E. Stratmann, 
 O. Yazyev, A. J. Austin, R. Cammi, C. Pomelli, J. W. Ochterski, 
 R. L. Martin, K. Morokuma, V. G. Zakrzewski, G. A. Voth, 
 P. Salvador, J. J. Dannenberg, S. Dapprich, A. D. Daniels, 
 O. Farkas, J. B. Foresman, J. V. Ortiz, J. Cioslowski, 
 and D. J. Fox, Gaussian, Inc., Wallingford CT, 2010.}.
All periodic solid calculations are performed in a modified version of \textsc{vasp 5.2}\cite{VASP1,VASP2,VASP3,VASP4,supplemental}.
All final computational results not directly reported in the paper, as well as
 all molecular and crystal structures used in this study, are compiled in the supplementary material\cite{supplemental}.

All \textsc{gaussian} calculations are performed using the 6-311++$G(3df,3pd)$ basis set,
 consistent with previous HSE studies\cite{HSE06}.
The exchange hole form of PBE\cite{PBE_hole} that is used within HSE is applied to all calculations, even when $\omega = 0$.
Default settings are used except for the fraction of non-convergent calculations, which we fix by switching
 to the quadratically-convergent self-consistent field (SCF) method.

All \textsc{vasp} calculations are performed using the manual-recommended PBE projector augmented wave (PAW)
 pseudopotentials\cite{VASP_PAW} with their default planewave basis sets,
 which are assumed to be transferrable to all DFT functionals considered in this paper.
The Brillouin zone integration is performed with the tetrahedron method on a $\Gamma$-centered grid, $12 \times 12 \times 12$ for cubic solids
 and $12 \times 12 \times 8$ for hexagonal solids.
The Fock exchange is down-sampled by half in each direction, consistent with convergence studies\cite{VASP_BZ}.
Fully sampled Fock exchange is applied as an additional perturbative correction to the band gap\cite{delta_hybrid}.
Minor modifications to the current version of \textsc{vasp} are required
 to calculate the 3 sX-LDA variants considered in this paper\cite{supplemental}.

Comparing DFT results to experiments may require finite temperature and quantum nuclear corrections
 that are not typically included within DFT itself.
The largest corrections occur for formation energies.
We account for these corrections at the level of G3 theory for the G3/99 test set\cite{G3_theory},
 but omit them for all other test sets.
As DFT becomes more accurate, the importance of accounting for these effects to correctly assess accuracy will grow.
Also, spin-orbit coupling is omitted from all calculations.
This can have a large effect on band gaps, but the SC/40 test set\cite{SC40} that we utilize
 has removed large spin-orbit effects from its experimental values.

Error distributions are quantified using the mean error (ME) with a (theory - experiment) sign convention,
 the mean absolute error (MAE), and the root-mean-square error (RMSE).
Figures show MAE, which is at present the preferred error average in quantum chemistry.

\section{Results and discussion}

The physically relevant region of the HSE parameter space is $0 \le \omega \le \infty$ and $0 \le a \le 1$.
For sampling and visualization purposes, the $\omega$ parameter is mapped to a bounded and dimensionless
range separation parameter,
\begin{equation}
 \widetilde{\omega} = \frac{2}{\pi} \arctan\left(\frac{\omega}{\omega_{\textrm{HSE06}}}\right),
\end{equation}
 with $\omega_{\textrm{HSE06}}$ = 0.208 \AA$^{-1}$ and $0 \le \widetilde{\omega} \le 1$.
This aligns the HSE06 parameterization to the center of the range separation axis, $\widetilde{\omega}_\textrm{HSE06} = 0.5$.
All points in HSE space are specified as $(\widetilde{\omega},a)$ ordered pairs.
In some cases, the range separation parameter is given as a screening length,
 which we derive using Eq. (\ref{rtf}) as $r_\mathrm{TF} = 0.58/\omega$.

A mixture of molecular and bulk solid tests are used to assess the accuracy of HSE functionals.
The SC/40 set\cite{SC40} contains 33 band gaps and 42 lattice constants of binary and elemental semiconductors
 (lattice constants of the 3 hexagonal structures are omitted from our study).
The G3/99 set\cite{G3_99} contains 223 formation energies, 86 IPs, and 58 EAs for small molecules
 (excited states of SH$_2^+$ and N$_2^+$ are omitted from our study).
The BH42/04 set\cite{BH42} contains 39 distinct barrier heights mostly for pairwise interactions between small molecules.
The T-96R set\cite{T96} contains 96 bond lengths of small molecules.
All of these test sets have been used before to assess the accuracy of HSE functionals\cite{HSE06,HSE_explore,SC40}.

\subsection{SC/40 band gaps\label{gap_section}}

\begin{table}
\caption{\label{gap_table}Comparison of $r_\mathrm{TF}$
 (from Eq. (\ref{rtf}), without semicore $d$-states in $\overline{\rho}$) and band gaps of the SC/40 set\cite{SC40}
 using 5 sX-DFT functionals described in the text, labelled by their exchange-correlation model, screening function, and screening length.}
\begin{tabular}{r | c | c c c c c | l}
 && LDA$_\mathrm{G}$ & LDA$_\mathrm{L}$ & LDA$_\mathrm{L}$ & PBE & PBE & Expt.\\
 &  & exp & exp & erfc & erfc & erfc & band\\
 & & $r_\mathrm{TF}$ & $r_\mathrm{TF}$ & $r_\mathrm{TF}$ & $r_\mathrm{TF}$ & $0.787$ & gap \\ 
 \cline{3-7}
 Solid& $r_\mathrm{TF} (\textrm{\AA})$ & \multicolumn{5}{c |}{sX-DFT band gap (eV)} & (eV) \\
\hline
C			& 0.389 & 5.19 & 4.44 & 4.27 & 4.37 & 5.45 & 5.48 \\
Si			& 0.479 & 1.34 & 0.74 & 0.64 & 0.73 & 1.08 & 1.17 \\
Ge			& 0.489 & 0.06 & 0.31 & 0.36 & 0.47 & 0.90 & 0.74 \\
SiC			& 0.430 & 2.52 & 1.62 & 1.47 & 1.56 & 2.25 & 2.42 \\
BN			& 0.391 & 6.01 & 4.82 & 4.62 & 4.77 & 6.14 & 6.22 \\
BP			& 0.438 & 2.07 & 1.41 & 1.30 & 1.37 & 1.88 & 2.4 \\
BAs			& 0.450 & 1.88 & 1.23 & 1.17 & 1.26 & 1.65 & 1.46 \\
AlN			& 0.430 & 5.29 & 4.59 & 4.31 & 4.47 & 5.92 & 6.13 \\
AlP			& 0.481 & 2.57 & 1.78 & 1.66 & 1.79 & 2.22 & 2.51 \\
AlAs			& 0.490 & 2.32 & 1.58 & 1.49 & 1.61 & 1.98 & 2.23 \\
AlSb			& 0.510 & 1.60 & 1.15 & 1.16 & 1.26 & 1.50 & 1.68 \\ 
GaN			& 0.441 & 2.39 & 1.90 & 1.68 & 1.83 & 3.63 & 3.50 \\
$\beta$-GaN	& 0.438 & 2.65 & 2.12 & 1.89 & 2.03 & 3.32 & 3.30 \\
GaP			& 0.480 & 1.93 & 1.78 & 1.66 & 1.82 & 2.21 & 2.35 \\
GaAs 		& 0.489 & 0.56 & 0.94 & 0.99 & 1.14 & 1.82 & 1.52 \\
GaSb 		& 0.508 & 0.00 & 0.52 & 0.61 & 0.71 & 1.07 & 0.73 \\
InN 			& 0.459 & 1.06 & 0.13 & 0.01 & 0.05 & 0.95 & 0.69 \\
InP 			& 0.498 & 1.35 & 0.91 & 0.83 & 1.01 & 1.56 & 1.42 \\
InAs 			& 0.506 & 0.42 & 0.11 & 0.07 & 0.22 & 0.73 & 0.41 \\
InSb 			& 0.524 & 0.23 & 0.16 & 0.17 & 0.28 & 0.70 & 0.23 \\
ZnS 			& 0.479 & 3.82 & 2.80 & 2.59 & 2.84 & 3.77 & 3.66 \\
ZnSe 		& 0.490 & 2.90 & 2.05 & 1.87 & 2.10 & 2.91 & 2.70 \\
ZnTe 		& 0.508 & 2.44 & 1.91 & 1.83 & 2.00 & 2.64 & 2.38 \\
CdS 			& 0.496 & 2.74 & 1.67 & 1.46 & 1.71 & 2.46 & 2.55 \\
CdSe 		& 0.506 & 2.13 & 1.21 & 1.03 & 1.26 & 1.92 & 1.90 \\
CdTe 		& 0.524 & 1.83 & 1.24 & 1.14 & 1.32 & 1.84 & 1.92 \\
MgO 		& 0.422 & 6.74 & 5.64 & 5.17 & 5.40 & 7.47 & 7.22 \\
MgS 			& 0.488 & 4.67 & 3.87 & 3.63 & 3.92 & 4.74 & 5.4 \\
MgSe 		& 0.478 & 2.78 & 2.10 & 1.95 & 2.06 & 2.70 & 2.47 \\
MgTe 		& 0.521 & 3.12 & 2.70 & 2.62 & 2.82 & 3.32 & 3.6 \\
BaS 			& 0.520 & 3.22 & 2.41 & 2.30 & 2.48 & 3.00 & 3.88 \\
BaSe 		& 0.528 & 2.93 & 2.16 & 2.05 & 2.21 & 2.66 & 3.58 \\
BaTe 		& 0.545 & 2.36 & 1.68 & 1.58 & 1.73 & 2.04 & 3.08 \\
\hline
ME 			&            &-0.24 &-0.83 &-0.95 &-0.80 &-0.08 & \\
MAE 		&            & 0.38 & 0.83 & 0.95 & 0.80 & 0.28 & \\
RMSE 		&            & 0.48 & 0.94 & 1.08 & 0.94 & 0.38 & \\
\end{tabular}
\end{table}

The comparison between sX-LDA and HSE is confined to the SC/40 band gaps
 because most sX-LDA studies have focused on band structures.
We connect sX-LDA to HSE through a sequence of four functional modifications
 with the results shown in Table \ref{gap_table}.
In sX-LDA, the long-range exchange energy is computed
 from integration of a globally-weighted local exchange energy density (LDA$_\textrm{G}$),
\begin{equation}
 E_\mathrm{x}^\mathrm{LDA_G,LR} = \int d\mathbf{r} \rho(\mathbf{r}) \epsilon_\mathrm{x}^\mathrm{LDA}[ \rho(\mathbf{r})] 
 f\left(\overline{\rho}^{1/3} r_\mathrm{TF}\right).
\end{equation}
The use of $\overline{\rho}$ instead of $\rho(\mathbf{r})$ in the weight function $f$ was
 initially chosen based on improved band gaps of silicon\cite{sX_LDA}.
$f$ is based on an exchange-hole formulation of LDA,
 and consistency with the exchange-hole formulation of PBE used in HSE\cite{PBE_hole}
 requires a local weighting (LDA$_\textrm{L}$),
\begin{equation}
 E_\mathrm{x}^\mathrm{LDA_L,LR} = \int d\mathbf{r} \rho(\mathbf{r}) \epsilon_\mathrm{x}^\mathrm{LDA}[ \rho(\mathbf{r})] 
 f\left(\rho(\mathbf{r})^{1/3} r_\mathrm{TF}\right).
\end{equation}
The effect of switching between LDA$_\mathrm{G}$ and LDA$_\mathrm{L}$ in sX-LDA is shown
 in the first two sX-DFT columns of Table \ref{gap_table}.
As in silicon, this change results in a systematic band gap underestimation in relation to the more accurate LDA$_\mathrm{G}$.
The following two modifications are replacement of the $\exp(-r/r_\mathrm{TF})$ screening function with an optimal $\mathrm{erfc}(\omega r)$
 approximant as in Eq. (\ref{screen_approx}) and the replacement of the LDA model of correlation and long-range exchange
 with the PBE model used by HSE.
As shown within the third and fourth sX-DFT columns of Table \ref{gap_table}, this produces small changes
 in the band gaps.
The final modification is to choose a single $\omega$ value for all materials to recover the HSE form.
The variations in $r_\mathrm{TF}$ appear to be small because of its sixth-root dependence on $1/\overline{\rho}$ in Eq. (\ref{rtf}).
As a result, a single optimized $\omega$ value is able to surpass the performance of sX-LDA 
 with material-dependent $r_\mathrm{TF}$ values for this test set.
We denote the optimized HSE approximation to sX-LDA as sX-PBE.

\begin{figure}
\includegraphics{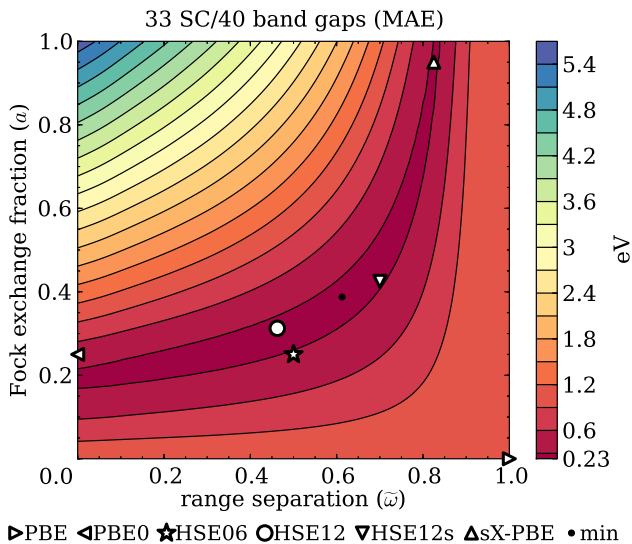}
\caption{\label{gaps}Semiconductor band gap errors over HSE space. A subset of functionals with approximately equivalent accuracy
 extend from (0,0.19) to (0.83,1).}
\end{figure}

The consistency of our results with previous sX-LDA results is mixed.
We agree with prior results\cite{GKS,sX_LDA,sX_Freeman,sX_Wang} on Si
 but produce significantly smaller band gaps for Ge and GaAs
 compared to other reported sX-LDA values\cite{GKS,sX_Freeman,sX_Wang,sX_Robertson}.
These discrepancies can be explained by the absence of valence-core interaction terms\cite{valence2core}
 in prior sX-LDA results.
The changes are small in Si but large in Ge and GaAs.
\textsc{vasp} includes the effect of these missing terms as all-electron corrections within atomic spheres\cite{VASP_hybrid}.

The MAE of SC/40 band gaps over the whole $(\widetilde{\omega},a)$ HSE space is plotted in Fig. (\ref{gaps}).
Errors near $(1,0)$ come from band gap underestimation in the PBE model that is
 largely attributed\cite{deriv_discon} to the absence of discontinuities in $\delta E_\mathrm{xc}^\mathrm{PBE} / \delta \rho(\mathbf{r})$.
Errors near $(0,1)$ come from band gap overestimation in Hartree-Fock theory\cite{HF_gap}
 (the additional correlation potential has relatively little effect here).
In a GKS theory context, the 5 eV error variation in going from $(1,0)$ to $(0,1)$
 originates from tuning the amount of nonlocality in the GKS potential
 to compensate for the lack of $\delta E_\mathrm{xc} / \delta \rho(\mathbf{r})$ discontinuities in the semilocal model.
Within the HSE space, there is a 1-dimensional subset of similarly good functionals for band gap estimation as judged by MAE.
This non-uniqueness is not surprising given that GKS theory introduces greater flexibility in functional form
 without adding a commensurate amount of new mathematical or physical constraints.
Requiring that GKS theory reproduce the exact 1-electron reduced density matrix (1RDM)
 instead of just electron density can achieve uniqueness\cite{1RDM_HK}.
However, this introduces the ongoing problem of constructing GKS functionals
 that produce the fractional occupation numbers characteristic of 1RDMs for interacting electrons, 
 even for pure states with an integer number of electrons\cite{GKS_occ}.

We can argue for the accuracy of band gaps in HSE based on its similarity to an electron quasiparticle theory.
Specifically, the simplification of Hedin's equations using the COHSEX approximation\cite{COHSEX} 
 results in the self-energy (assuming a natural spin-dependent extension)
\begin{align}\label{COHSEX}
 \Sigma_{\sigma,\sigma'}^{\mathrm{COHSEX}} &(\mathbf{r},\mathbf{r}') = - \rho_{\sigma,\sigma'}(\mathbf{r},\mathbf{r}') W_{\sigma,\sigma'}(\mathbf{r},\mathbf{r}') \notag \\
  + \frac{1}{2} & \delta_{\sigma,\sigma'} \delta(\mathbf{r}-\mathbf{r}') \left( W_{\sigma,\sigma'}(\mathbf{r},\mathbf{r}') - \frac{1}{|\mathbf{r}-\mathbf{r}'|}\right)
\end{align}
 with a non-local sX operator and a local potential derived from the Coulomb hole, both
 determined by a statically screened Coulomb interaction $W_{\sigma,\sigma'}(\mathbf{r},\mathbf{r}')$.
Minimal, yet realistic, semiconductor screening models\cite{eps_model} contain the bulk static dielectric constant $\varepsilon_0$ 
 and a Thomas-Fermi-like screening length $r_\mathrm{TF}$.
A simplified example is
\begin{equation}\label{eps}
 W_{\sigma,\sigma'}(\mathbf{r},\mathbf{r}') \approx \frac{\varepsilon^{-1}_0 + (1-\varepsilon^{-1}_0) \exp(-|\mathbf{r}-\mathbf{r}'|/r_\mathrm{TF})}{|\mathbf{r}-\mathbf{r}'|},
\end{equation}
 which has a constant Coulomb hole, $(\varepsilon_0^{-1}-1)/(2 r_\mathrm{TF})$,
 corresponding to the polarization energy of an electron in a spherical cavity of radius $r_\mathrm{TF}$
 within a dielectric of permittivity $\varepsilon_0$\cite{Jost}.
We presume that oversimplification of $W$ breaks the connection between sX and the Coulomb hole,
 the latter being reasonably approximated by the PBE local potential that 
 corresponds to correlation and long-range exchange.

The position and shape of the 1-dimensional subset of accurate functionals within Fig. (\ref{gaps})
 can be explained by the connection between HSE and COHSEX.
For any one semiconductor, we have one constraint, the band gap, and two free parameters, $(\widetilde{\omega},a)$,
 therefore each can be fit perfectly along a line in HSE space.
The near-overlap of all these lines results from similar parameters for all members of the SC/40 set.
The effective $W$ of HSE,
\begin{equation}\label{hseW}
 W^\mathrm{HSE}_{\sigma,\sigma'} (\mathbf{r},\mathbf{r}') = a \frac{\mathrm{erfc}(\omega |\mathbf{r}-\mathbf{r}'| )}{|\mathbf{r}-\mathbf{r}'|},
\end{equation}
 can fit Eq. (\ref{eps}) when $r_\mathrm{TF}^{-1} = 0$ or $\varepsilon_0^{-1} = 0$.
We assume similar gap-invariant variations of $(r_\mathrm{TF}^{-1},\varepsilon_0^{-1})$ in Eq. (\ref{eps})
 from reasonable physical values to $(r_\mathrm{TF}^{-1} = 0, \widetilde{\varepsilon}_0^{-1} > \varepsilon_0^{-1})$ or
 $(\widetilde{r}_\mathrm{TF}^{-1} < r_\mathrm{TF}^{-1},\varepsilon_0^{-1}=0)$.
This explains the optimal $a \approx \widetilde{\varepsilon}_0^{-1} \approx 0.2$ that is larger than
 the average value for the SC/40 set, $\varepsilon_0^{-1} \approx 0.1$.
It also explains the behavior in Table \ref{gap_table} of an optimal sX-PBE screening length
 that is longer than the Thomas-Fermi screening lengths.
This malleability also explains how the same arguments can be used to support both hybrid\cite{PBE_eps}
 and sX\cite{GKS} DFT functionals, with comparable validity and empirical success.

\subsection{G3/99 formation energies\label{form_section}}

\begin{figure}
\includegraphics{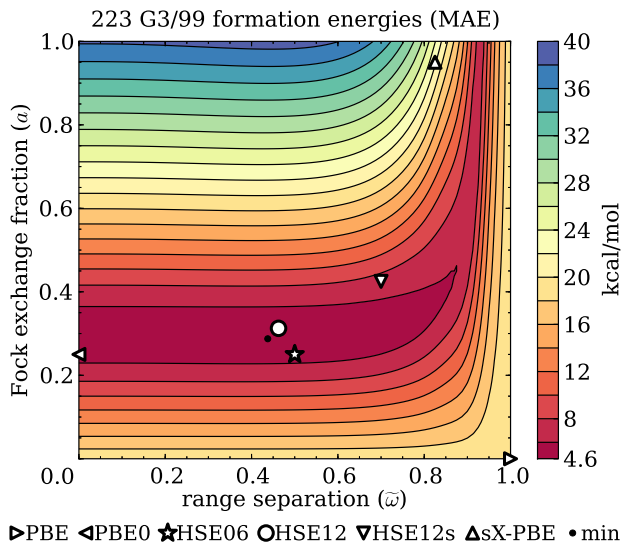}
\caption{\label{formation}Molecular formation energy errors over HSE space. A subset of functionals
 with approximately equivalent accuracy extend from (0,0.3) to (0.93,1). }
\end{figure}

The MAE of G3/99 formation energies over HSE space is shown in Fig. (\ref{formation}).
As with the SC/40 band gaps, there is a 1-parameter family of similarly accurate functionals,
 here with a larger exchange fraction ($30\%$ versus $20\%$) in the hybrid limit
 and a smaller effective screening length ($0.33$ \AA \  versus $0.79$ \AA) in the sX limit.
A negligible $\widetilde{\omega}$-dependence for screening lengths larger than $2.4$ \AA \ ($\widetilde{\omega} \le 0.55$)
 shows that the PBE model of long-range exchange performs well beyond that length scale,
 at least within small molecules.
 GKS orbitals are only weakly dependent on details of the exchange functional\cite{back_action,Ctail},
 therefore errors in the exchange functional itself are the main source of MAE variation.
In particular, the largest errors occur near $(0,1)$, where the exact exchange energy is used.
This is an example of error cancellation between exchange and correlation models
 that is often observed in DFT\cite{xc_cancellation}.

The errors in formation energy can be attributed to a systematic bias.
Similar to the results for band gaps, the formation energy is underestimated near $(1,0)$ in Fig. (\ref{formation})
 and overestimated near $(0,1)$, with ME passing through zero near the region of minimal MAE.
Here, we use the sign convention of positive formation energies for stable molecules.
We observe that the PBE exchange energy overestimates the Fock exchange energy,
 $E_\mathrm{x}^\mathrm{PBE} > E_\mathrm{x}^\mathrm{HF}$,
 with larger errors for molecules than for atoms.
Also, we observe that the PBE correlation energy underestimates
 the experimental correlation energy, $E_\mathrm{c}^\mathrm{PBE} \lesssim E_\mathrm{c}^\mathrm{exp}$,
 with errors again larger for molecules than for atoms.
These observations are consistent with a cancellation of errors between
 exchange and correlation contributions to the formation energy.

The similar error trends in Figs. (\ref{gaps}) and (\ref{formation}) suggest that HSE functionals are
 ``right for the right reasons''.
The ability of HSE functionals to estimate band gaps is explained by quasiparticle theory,
 which can be extended to cover total energy approximation.
This is related to ongoing efforts\cite{Sohrab} to compute accurate total energies from self-consistent 1-electron Green's functions.
Specifically, the HSE functional can be interpreted as the Galitskii-Migdal formula\cite{green_energy}
 applied to the frequency-independent COHSEX self-energy in Eq. (\ref{COHSEX}) to calculate an exchange-correlation energy,
\begin{equation}\label{sigma_xc}
 E_\mathrm{xc} =  \frac{1}{2}\sum_{\sigma,\sigma'} \int d\mathbf{r} d\mathbf{r}'  \rho_{\sigma',\sigma}(\mathbf{r}',\mathbf{r}) \Sigma_{\sigma,\sigma'}(\mathbf{r},\mathbf{r}').
\end{equation}
The effective HSE self-energy is particularly suited for total energy calculations
 since it is constrained to exactly reproduce the total energy of the uniform electron gas and PBE gradient corrections thereof.

\subsection{G3/99 ionization potentials and electron affinities\label{charged_up}}

\begin{figure*}
\includegraphics{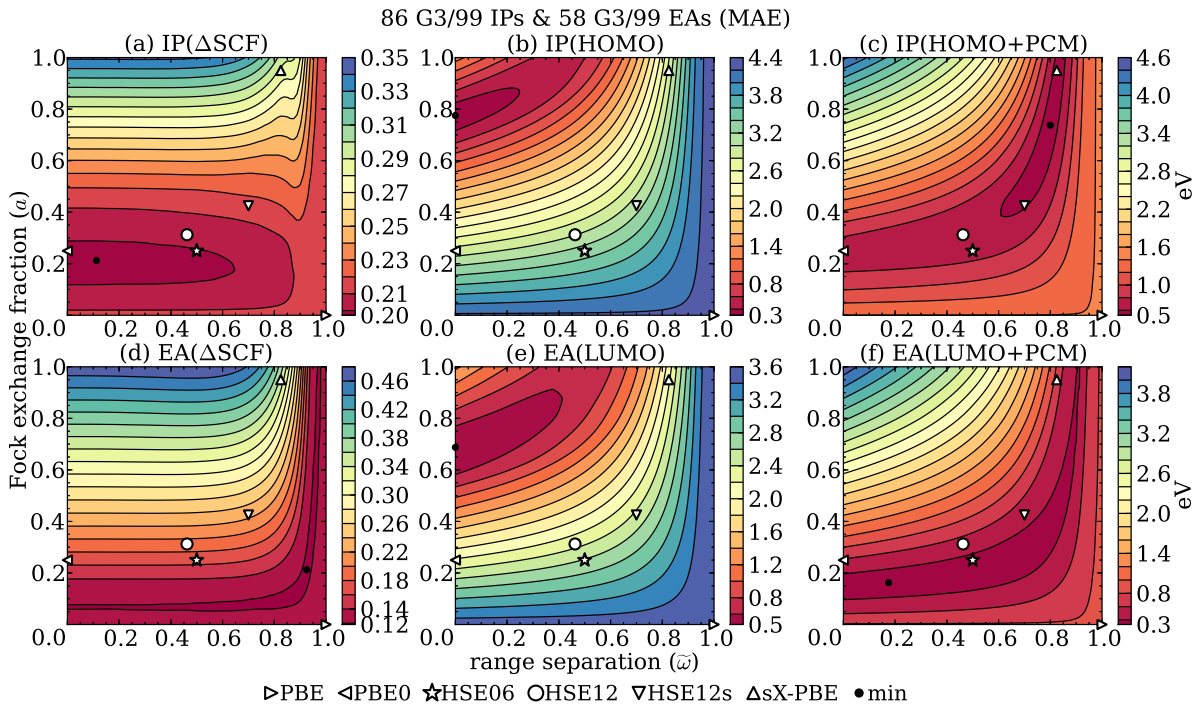}
\caption{\label{exciting}IP and EA errors in HSE space, comparing $\Delta$SCF
 to eigenvalue-based estimates and including polarization corrections.}
\end{figure*}

Vertical IPs and EAs can be computed with a GKS functional using either total energy differences
 between charge states ($\Delta$SCF) or the HOMO/LUMO eigenvalues, if the functional is accurate
 and free of discontinuities in $\delta E_\mathrm{xc}/ \delta \rho(\mathbf{r})$.
The G3/99 test set contains adiabatic IPs and EAs, which requires the use of relaxed geometries
 in $\Delta$SCF calculations and a relaxation energy correction for eigenvalue-based vertical excitation energies
(the total energy difference of the neutral molecule between neutral and ionized geometries).
$\Delta$SCF IPs and EAs are shown in Figs. (\ref{exciting}a) and (\ref{exciting}d).
The variation of errors is smaller than in the formation energies of Fig. (\ref{formation}),
 which results from increased cancellation of errors because
 two charge states of a molecule are more similar than a molecule and its dissociated atoms.
There is still the same subset-of-accurate-functionals trend, but it
 does not stand out as strongly without a large background of systematic errors.
The eigenvalue-based IP and EA estimates in Figs. (\ref{exciting}b) and (\ref{exciting}e)
 show much greater variation over HSE space, comparable to the band gaps in Fig. (\ref{gaps}).
There is still a subset of accurate functionals, but it is shifted to large Fock exchange fractions ($a \ge 70\%$)
 that are incompatible with accurate formation energies or band gaps.

\begin{figure}
\includegraphics{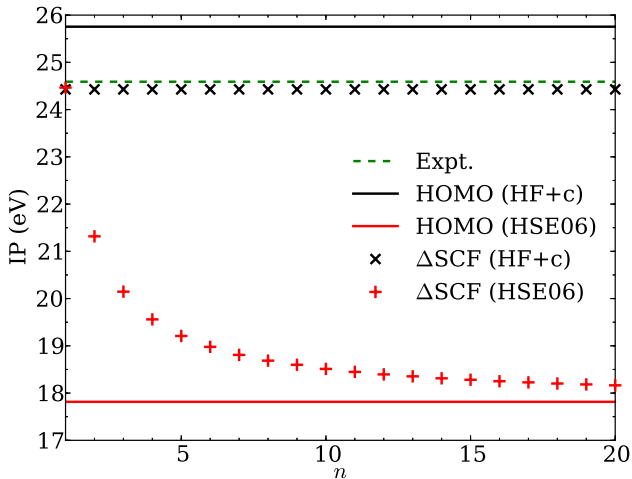}
\caption{\label{loca}IP of a dilute He$_n$ cluster, comparing $-\epsilon_\mathrm{HOMO}$(He$_n$) to $E$(He$_n^+$)$-E$(He$_n$)
using the HSE06 functional and full Fock exchange with PBE correlation (HF+c),
which correspond to the points $(0.5,0.25)$ and $(0.0,1.0)$ in HSE space.
In HSE06, the $\Delta$SCF and eigenvalue-based results converge for $n \rightarrow \infty$.
The electron-hole of He$_n^+$ is localized on one atom with HF+c and uniformly divided among all  $n$ atoms with HSE06.
Any $n$-dependence is a size-consistency error.}
\end{figure}

The discrepancy between $\Delta$SCF and eigenvalue-based IPs and EAs constitutes both delocalization error\cite{delocal}
 and size-consistency error\cite{size_consistency}.
Compared to $\Delta$SCF values,
 the eigenvalues underestimate IP and overestimate EA near $(1,0)$ in HSE space and vice versa near $(0,1)$.
If a hole or electron is added to a dilute gas of identical molecules,
 the charge will delocalize to avoid spurious Coulomb self-interaction and approach the erroneous eigenvalue-based IP/EA
 for functionals near $(1,0)$.
The opposite effect, localization of charge onto a single molecule, occurs near $(0,1)$
 and forces IP/EA to remain at the more accurate $\Delta$SCF value.
This dichotomy is demonstrated in Fig. (\ref{loca}) for a dilute He$_n$ gas with $n = 1,\cdots,20$.
DFT functionals with delocalization behavior are not size consistent,
 and making use of the $\Delta$SCF accuracy requires a single well-defined charge center
 such as a small molecule or a point defect inside a crystal\cite{Peter}.
The subset of functionals that are accurate for formation energies and band gaps
 is well within the delocalization regime and uniformly lacks size consistency.
In the localization regime, the eigenvalues are no longer equivalent to charge excitations
 because the localized charges produces nonlinear corrections.

Considering IP as an example, we can account for the difference between
 $\Delta$SCF and eigenvalue-based estimates with a continued appeal to quasiparticle theory,
 building on the arguments for band gap accuracy in section \ref{gap_section}
 and total energy accuracy in section \ref{form_section}.
For a system $X$ and a DFT functional with a COHSEX-like form, the IP discrepancy can be written as
\begin{align}\label{koopish}
 E(X^+) - E(X) = -\epsilon_\mathrm{HOMO}(X) - \Delta_\mathrm{relax} \notag \\+ \frac{1}{2}\sum_{\sigma,\sigma'}\int d\mathbf{r} d\mathbf{r}'
 \left(\frac{1}{|\mathbf{r}-\mathbf{r'}|}-W_{\sigma,\sigma'}(\mathbf{r},\mathbf{r'})\right) \rho_\sigma^+(\mathbf{r}) \rho_{\sigma'}^+(\mathbf{r}')
\end{align}
 by separating orbital relaxation effects $\Delta_\mathrm{relax}$ from terms that assume the cation to have
 the same $W$, orbitals, and energies as the neutral system.
Within Hartree-Fock theory, the last term vanishes and this is the essential content of Koopmans' theorem\cite{koopman},
 which guarantees that $\Delta$SCF doesn't produce a larger IP
 than $-\epsilon_\mathrm{HOMO}$ since $\Delta_\mathrm{relax} \ge 0$.
The last term is a polarization correction to the self-interaction of the hole charge $\rho^+$
 and, in an accurate quasiparticle theory, it should cancel $\Delta_\mathrm{relax}$
 to produce consistency between IP estimates.
If we assume this to be true for a COHSEX model with a realistic $W$,
 then errors in HSE can be attributed to the lack of an unscreened Coulomb tail,
 $W\rightarrow 1/|\mathbf{r}-\mathbf{r}'|$, extending out from a small molecule.
The net effect is over-screening in small molecules compared to semiconductors,
 which is consistent with the offset between Figs. (\ref{gaps}) and (\ref{exciting}b).

The difference between $1/|\mathbf{r}-\mathbf{r}'|$ and $W$ must originate from an electronically polarizable substance.
Since $W_\mathrm{HSE}$ remains screened even in vacuum,
 the effect there can be assigned to unphysical vacuum polarization.
This error cannot be removed without environment-dependence and inhomogeneity in $W_\mathrm{HSE}$,
 but we can model it by adding comparable errors into $\Delta$SCF calculations.
Specifically, we introduce an artificial vacuum polarization into total energies
 with the polarizable continuum model (PCM)\cite{PCM}.
Over-screening errors of eigenvalue-based IPs and EAs are corrected by subtracting out the errors
 modeled by the difference between $\Delta$SCF+PCM and $\Delta$SCF values.
The results in Figs. (\ref{exciting}c) and (\ref{exciting}f) are less accurate than $\Delta$SCF values,
 but they succeed in recovering previous error trends in HSE space.
Similar over-screening errors should also occur for localized defect states in crystals,
 but with the discrepancy in polarization reduced by $\varepsilon_0^{-1}$.
Another solution is to include long-range Fock exchange to model the correct asymptotic of $W$,
 with its range parameters tuned to model the size of the molecule and approximately satisfy Koopmans' theorem\cite{leeor}.
Given our results with HSE, caution must be exercised when tuning range-separated hybrid functionals
 into a regime where HOMO and $\Delta$SCF values of the IP match,
 because this may degrade the accuracy of other physical properties.

\subsection{Conformational tests: BH42/04 barrier heights, T-96R bond lengths, and SC/40 lattice constants}

\begin{figure*}
\includegraphics{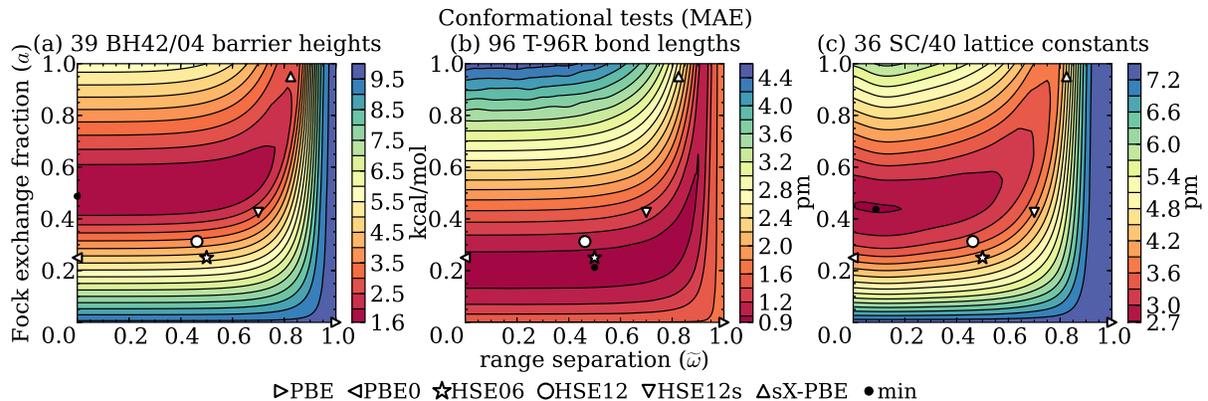}
\caption{\label{conform}Binary reaction barrier height, molecular bond length, and semiconductor lattice constant errors in HSE space.}
\end{figure*}

The final set of tests are based on atomic conformations that are less pronounced than full formation,
 which leads to less variation of error in HSE space than the formation energies in Fig. (\ref{formation}).
For the barrier heights in Fig. (\ref{conform}a),
 the optimal $\omega$ value at $a=1$ is consistent with previous results\cite{HSE_explore}.
The error trend in molecular bond lengths in Fig. (\ref{conform}b) matches well with
 molecular formation energies in Fig. (\ref{formation}).
The increased $\widetilde{\omega}$-dependence at small $\widetilde{\omega}$ and large $a$
 of the lattice constants in Fig. (\ref{conform}c)
 are the result of unconverged Brillouin zone sampling
 of the long-range exchange tail.
In all cases, the minimum in MAE closely corresponds to the zero of ME.
The net effect is to reduce small systematic errors in PBE that overestimate bond lengths and lattice constants
 and underestimate barrier heights.

\subsection{Outlook on method development}

\begin{figure}
\includegraphics{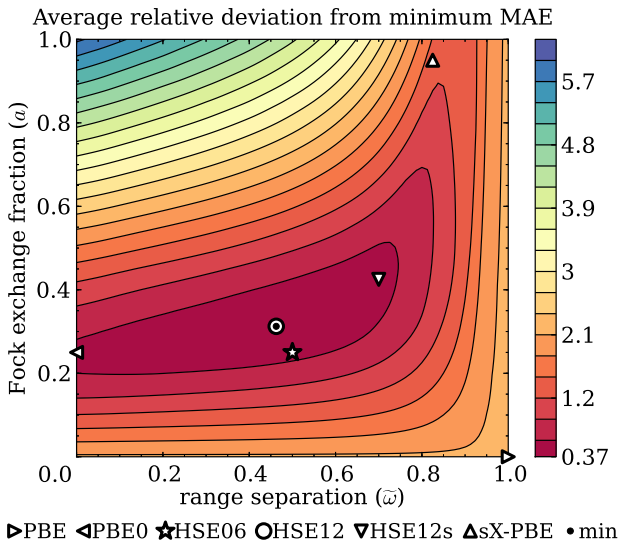}
\caption{\label{best}Aggregate error in HSE space. The minimum of this error metric occurs at the HSE12 parameterization.
 HSE06 and HSE12s lie on a contour of equal error. }
\end{figure}

Our combined computational survey and analysis of HSE functional space suggests several new directions for method development
 with varying degrees of expected difficulty and efficacy.
We have shown the approximate correspondence of sX-LDA to the HSE functional form.
Given the limited performance of sX-LDA for properties other than band gaps\cite{sX_correction},
 we conclude that HSE is both conceptually and empirically superior because of its use of PBE instead of LDA
 for correlation and long-range exchange.
Any benefit to the system-dependence of the sX-LDA screening length that is negligible for solids and ill-defined for molecules
 is offset by optimization of a system-independent HSE screening length.
Within the HSE form, we can recommend new parameters that are modest improvements over HSE06.
We base this on an aggregate error metric shown in Fig. (\ref{best}),
\begin{equation}
 \frac{1}{\mathrm{\# \ of \ tests}}\sum_{i}^\mathrm{tests} \left(\frac{\mathrm{MAE}_i(\widetilde{\omega},a)}{\min_{\widetilde{\omega}',a'} \mathrm{MAE}_i(\widetilde{\omega}',a')} - 1 \right).
\end{equation}
The eigenvalue-based tests are excluded because of their systematic bias.
This metric puts emphasis on the more sensitive tests and is zero if the MAE of each test can be simultaneously minimized.
The ``HSE12'' functional, with $\omega=0.185$ \AA$^{-1}$ and $a=0.313$,
 optimizes accuracy by minimizing the metric.
The ``HSE12s'' functional, with $\omega=0.408$ \AA$^{-1}$ and $a=0.425$,
 minimizes screening length while preserving the accuracy of HSE06.
HSE12s has half the screening length of HSE06, which can reduce computational costs
 in implementations that make use of spatial decay in short-range Fock exchange.
A summary of functionals and their performance is given in Table \ref{summary}.

The reduced cost of HSE12s can be demonstrated by the relative speed-up of an example: bulk aluminum.
In $\textsc{gaussian}$ (using the basis from Ref. 51), the relative speed of HSE12s compared to HSE06 is $1.6$.
The cause of this improvement is exchange integral screening \cite{HSE_screen}.
In $\textsc{vasp}$, the speed-up is $1.4$.
Here, the improvement comes from reduced Brillouin zone sampling of the sX term for HSE12s compared to HSE06 \cite{VASP_BZ},
 requiring $5 \times 5 \times 5$ points instead of $6 \times 6 \times 6$
 to achieve equivalent convergence.

Further progress in increasing functional accuracy will require alterations of the HSE form.
More complicated screening forms have been proposed\cite{HISS},
 which enable more realistic $W$ models, such as in Eq. (\ref{eps}),
 and may help to align the optimal choice of screening for multiple physical properties.
The PBE model was designed for exclusively semilocal treatment of exchange,
 and a recent result\cite{PBEmol} shows the benefits of refitting PBE dependent on how
 exchange is split into Fock and semilocal components,
 because it enables a fit of PBE to the residual, rather than total, exchange energy.
Based on the strong ties between HSE and COHSEX outlined in this paper,
 it may be worthwhile to pursue semilocal DFT functionals based on a sX/Coulomb-hole partition\cite{AEM}
 instead of an exchange/correlation partition.
Given the observed error trends in HSE space, it may be possible to construct a semilocal sX functional
 by taking the subset of accurate functionals to the $(\omega,a) \rightarrow \infty$ limit.
Unfortunately, these suggestions are not likely to fix the large errors in HSE HOMO/LUMO eigenvalues for molecules,
 because they do not introduce any nonlocal screening effects.

The inevitability of nonlocality can be argued in the context of GKS theory as an approximation to COHSEX theory.
The screened exchange operator in Eq. (\ref{COHSEX}) is a direct product of $\rho$ and $W$,
 which is a spatially localized operator for basic electronic phases.
In an insulator\cite{nearsight}, $\rho$ decays exponentially with increasing $|\mathbf{r}-\mathbf{r}'|$,
 while $W$ decays only algebraically.
In a metal, it is $\rho$ that decays algebraically and $W$ that decays exponentially.
In either case, the product is localized by one constituent while the other builds
 nonlocal dependence into the quantitative details.
Hypothetical examples of nonlocal dependence are electron interference in $\rho$ resulting from
 proximity to a scatterer in a metal or increased screening in $W$ from proximity to a highly polarizable object in an insulator.
In the context of DFT, the exchange-correlation hole is observed to be spatially localized,
 but no limits have been placed on its sensitivity to distant perturbations\cite{farsight}.

DFT modeling can be roughly split between quantities that are tractable (albeit expensive)
 to compute exactly (e.g. Fock exchange energy) and those that cannot be generally computed (e.g. exact correlation energy).
At the level of COHSEX theory, $W$ is contained within the first category
 and can be explicitly constructed from an independent-electron response function,
\begin{align}\label{trueW}
 \chi_{\sigma,\sigma'}^0(\mathbf{r},\mathbf{r}') &= \sum_{i,j} \frac{f_i - f_j}{\epsilon_i - \epsilon_j}\psi^*_{i,\sigma}(\mathbf{r})\psi_{j,\sigma}(\mathbf{r})\psi^*_{j,\sigma'}(\mathbf{r}')\psi_{i,\sigma'}(\mathbf{r}') \notag \\
 W^{-1}_{\sigma,\sigma'}(\mathbf{r},\mathbf{r}') &= \left[ \frac{1}{|\mathbf{r}-\mathbf{r}'|} \right]^{-1} - \chi_{\sigma,\sigma'}^0(\mathbf{r},\mathbf{r}'),
\end{align}
 for orbitals $\psi_i$, energies $\epsilon_i$, and occupations $f_i$, and where `$^{-1}$' refers to operator inversion.
Using such forms for $W$, promising results for total energies and charge excitations
 have been demonstrated for self-consistent quasiparticle methods\cite{scgw1,scgw2,scgw3}
 and much of this accuracy is preserved by the COHSEX approximation\cite{scCOHSEX}.
It is conceivable that the precision and reliability necessary to attain the goal of ``chemical accuracy''
 will only be achieved with a detailed treatment of $W$ in the same way orbital-free DFT
 has not been able to achieve a level of accuracy comparable to methods containing the details of electronic orbitals\cite{orbital_free}.
The construction of $\rho$ from electronic orbitals,
\begin{equation}
\rho_{\sigma,\sigma'}(\mathbf{r},\mathbf{r}') = \sum_{i} f_i \psi_{i,\sigma}(\mathbf{r}) \psi^*_{i,\sigma'}(\mathbf{r}'),
\end{equation}
 contains its essential nonlocal character and accounts for much of the success of Kohn-Sham DFT.
In principle, $\rho$ and $W$ have equal importance in COHSEX theory, and putting this into practice
 in COHSEX-inspired DFT functionals will require the use of Eq. (\ref{trueW}).

The form of $W$ in Eq. (\ref{trueW}) highlights an open problem in self-consistent quasiparticle theory.
Quasiparticle and GKS methods are designed to model charge excitations,
 but the electronic polarizability represented in Eq. (\ref{trueW}) is supposed to arise from neutral excitations.
The use of quasiparticle energies in Eq. (\ref{trueW}) 
produces a systematic overestimate of band gaps,
 which has been removed with vertex corrections encoding electron-hole interactions\cite{GWvertex}.
Otherwise, the non-self-consistent evaluation of Eq. (\ref{trueW}) using DFT methods is found to produce
 accurate mean-field models of static polarization\cite{GW0}.
Similar observations have been made for HSE:
 that it approximates the optical gap rather than the charge gap
 when there is a significant difference between the two\cite{HSEopt},
 even though it is a GKS theory that should be approximating the charge gap.
A heuristic explanation for this effect is that the universal lack of exchange screening
 in HSE can be attributed to electrons (holes) being screened by both a ground-state polarization and
 an excess hole (electron).
Just as with the excess polarization argument in section \ref{charged_up},
 this energy can be removed from the eigenvalues to improve their correspondence to quasiparticles,
 and in this case it can be re-identified for $\epsilon_\mathrm{LUMO}^\mathrm{HSE} - \epsilon_\mathrm{HOMO}^\mathrm{HSE}$
 as the Coulomb binding energy between an electron and hole.
It may be possible for a self-consistent COHSEX-based theory to avoid vertex corrections
 if it can define a consistent pair of mean fields: 
 a polarization mean field, meant to model neutral excitations for Eq. (\ref{trueW}),
 and a quasiparticle mean field containing the COHSEX self-energy from Eq. (\ref{COHSEX}).
Based on existing theory, it is possible to construct GKS and optimized effective potential (OEP)\cite{OEP}
 methods from the same COHSEX-based total energy functional,
 with the GKS form serving as the quasiparticle theory and the OEP form serving as the polarization model.
However, a consistent theoretical framework for such a ``double'' mean-field theory does not yet exist.
The theory would need to specify the interrelations between the orbitals and energies of the two mean fields
 and how they should be used in concert to calculate a total electronic energy.

\begin{table}
\caption{\label{summary}Summary of important functionals in HSE space and their performance
 on various tests discussed in this paper.}
\begin{tabular}{l r | r r r r r}
 Tests & 			& PBE	& PBE0 	& HSE06 	& HSE12 	& HSE12s \\
\hline
SC/40 &  ME 		& -1.13 	& 0.39 	& -0.24	& 0.06 	& -0.16\\
band & MAE 		& 1.13 	& 0.44 	& 0.32	& 0.23 	& 0.28\\
gaps & RMSE 		& 1.25 	& 0.49 	& 0.41	& 0.31 	& 0.36\\
(eV) & max 		& -2.46 	& 1.00 	& -0.90	& -0.66	& -0.95\\
\hline
G3/99 & ME 		& -19.0 	& -2.6 	& -2.6	& 1.4 	& 6.2\\
formation & MAE 	& 19.6 	& 5.2 	& 5.2		& 4.8 	& 7.3\\
energies & RMSE 	& 24.0 	& 7.4 	& 7.4		& 6.7		& 10.1\\
(kcal/mol) & max 	& -72.0 	& -29.8 	& -30.0	& 26.3 	& 40.7\\
\hline
G3/99 & ME 		& -0.07 	& -0.05 	& -0.05	& -0.05	& -0.06\\
ionization & MAE 	& 0.21 	& 0.20 	& 0.20 	& 0.20	& 0.22\\
potentials & RMSE 	& 0.28 	& 0.28 	& 0.29 	& 0.30	& 0.32\\
(eV) & max 		& 1.13 	& 1.72 	& 1.71 	& 1.86	& 2.01\\
\hline
G3/99 & ME 		& 0.08 	& -0.01 	& -0.02	& -0.04	& -0.07\\
electron & MAE 	& 0.13 	& 0.17 	& 0.17	& 0.19	& 0.22\\
affinities & RMSE 	& 0.18 	& 0.23 	& 0.23	& 0.26	& 0.30\\
(eV) & max 		& 0.78 	& 1.08 	& 1.07	& 1.14	& 1.20\\
\hline
BH42/04 & ME 		& -9.6 	& -4.6 	& -4.8 	& -3.6	& -2.9\\
barrier & MAE 		& 9.6 	& 4.6 	& 4.8 	& 3.7		& 2.9\\
heights & RMSE 	& 10.3 	& 4.9 	& 5.0 	& 3.9		& 3.2\\
(kcal/mol) & max 	& -20.0 	& -7.5 	& -7.7 	& -6.4	& -5.9\\
\hline
T-96R & ME		& 1.6 	& 0.0		& 0.0		& -0.4	& -0.7\\
bond & MAE 		& 1.6 	& 1.0		& 0.9		& 1.1		& 1.3\\
lengths & RMSE 	& 1.9 	& 1.5		& 1.4 	& 1.7		& 1.9\\
 (pm) & max 		& 5.5 	& 6.5		& 5.6 	& 7.9		& -7.1\\
\hline
SC/40 & ME 		& 7.4 	& 3.1 	& 3.5 	& 2.6		& 2.9\\
lattice & MAE 		& 7.4 	& 3.7 	& 4.1 	& 3.5		& 3.9\\
constants & RMSE 	& 8.5 	& 4.5 	& 4.9 	& 4.2		& 4.7\\
(pm) & max 		& 16.2 	& 10.1 	& 10.8	& 9.5		& 10.4\\
\end{tabular}
\end{table}

\section{Conclusions}

The novelty of our study comes from HSE's precarious perch between empirical and non-empirical DFT,
 with just two tunable parameters.
This enables a thorough sampling of parameter space, similar to recent studies on the PBE functional\cite{PBEtune}.
Besides the immediate benefit of further fine-tuning accuracy, the study reveals error trends
 supporting a connection between the HSE form and quasiparticle theory,
 which explains its success in approximating semiconductor band gaps.
The degree to which multiple physical properties can be simultaneously optimized
 by a common set of parameters is pleasantly surprising.
Unpleasant and unsurprising\cite{sX_correction} are the large discrepancies
 between $\Delta$SCF and eigenvalue estimates of small molecule IPs and EAs,
 which we argue to originate from the absence of environment-dependent screening
 in the HSE functional resulting from discrepancies
 between $W^\mathrm{HSE}$ in Eq. (\ref{hseW}) and $W$ in Eq. (\ref{trueW}).
It is encouraging that this error can be fixed by correcting the long-range tail of $W$\cite{leeor}.
However, this simple fix is not transferrable to more complicated systems such as a molecule near a metal surface,
 where a model $W$ such as Eq. (\ref{eps}) is not flexible enough to set $\varepsilon_0^{-1} = 0$ for the metal half-space
 and $\varepsilon_0^{-1} = 1$ for the empty half-space.
There is much to be gained from incorporating more sophisticated models of $W$
 into the screened Fock exchange component of GKS functionals,
 such as Eq. (\ref{trueW}) or a reasonable facsimile thereof\cite{continuumRPA}.

\begin{acknowledgments}
Sandia National Laboratories is a multi-program laboratory managed and  
operated by Sandia Corporation, a wholly owned subsidiary of Lockheed  
Martin Corporation, for the U.S. Department of Energy's National  
Nuclear Security Administration under contract DE-AC04-94AL85000.

JRC and JEM wish to acknowledge support from the National Science Foundation
 under grants No. DMR-0941645 and OCI-1047997
 and the Department of Energy under grant No. DE-FG02-06ER46286.
The computational resources used for this work were provided by the National Energy Research Scientific Computing Center (NERSC).

JEM thanks Norm Tubman, Leeor Kronik, and John Aidun for useful discussions related to this work.
\end{acknowledgments}


\begin{thebibliography}{69}
\bibitem [{PCM()}]{PCM}%
  \BibitemOpen
  \href@noop {} {}\bibinfo {note} {We use the default PCM implementation in
  \textsc{gaussian} 09, with the polarization cavity defined by the
  intersection of atomic spheres with van der Waals radii. The dielectric model
  is set to $\varepsilon_0 = \varepsilon_\infty \rightarrow \infty$. PCM
  calculations are performed with the PBE functional and results are assumed to
  be transferrable to all HSE functionals.}\BibitemShut {Stop}%
\end{thebibliography}
\end{document}